\newcommand \be{\begin{equation}}
\newcommand \ba{\begin{eqnarray}}
\newcommand \ea{\end{eqnarray}}
\newcommand \ee{\end{equation}}
\documentclass[twocolumn,showpacs,preprintnumbers,amsmath,amssymb]{revtex4}
\usepackage{epsfig}
\usepackage{color}
\begin{document}
\title{Fluctuations of the partial filling factors in competitive RSA from binary mixtures}
\author{Arsen V. Subashiev and Serge Luryi}
\affiliation{Department of Electrical and Computer Engineering,
State University of New York at Stony Brook, Stony Brook, NY,
11794-2350 }
\email[]{subashiev@ece.sunysb.edu}

\begin{abstract} Competitive random sequential adsorption on a line
from a binary mix of incident particles is studied using both an
analytic recursive approach and Monte Carlo simulations. We find a
strong correlation between the small and the large particle
distributions so that while both partial contributions to the fill
factor fluctuate widely, the variance of the total fill factor
remains relatively small. The variances of partial contributions
themselves are quite different between the smaller and the larger
particles, with the larger particle distribution being more
correlated. The disparity in fluctuations of partial fill factors
increases with the particle size ratio. The additional variance in
the partial contribution of smaller particle originates from the
fluctuations in the size of gaps between larger particles. We
discuss the implications of our results to semiconductor high-energy
gamma detectors where the detector energy resolution is controlled
by correlations in the cascade energy branching process.
\end{abstract}
 \maketitle

 \nopagebreak
 \baselineskip=20pt

\section{Introduction}

One dimensional irreversible random sequential adsorption (RSA)
has been of interest for several decades. Its numerous extensions
include RSA with particles expanding in the adsorption process
\cite{Rodg,viot,assl}, two-size particle adsorption
\cite{bartelt,hassan1,hassan2,Araujo}, and also RSA with an
arbitrary particle-size distribution function \cite{mao}. The
interest is due to the relevance of this process to a number of
physical phenomena in different fields of application, such as
information processing \cite{coff}, particle branching in impact
ionization \cite{ino} and crack formations in crystals under
external stress \cite{crack}. The simplest example of RSA is the
so-called {\it car parking problem} (CPP). In the context of CPP,
one studies the average number of particles (``cars'') adsorbed on
a long line and the variance of this number. Equivalently, one is
concerned with the distribution function for the size of gaps
between the parked cars (see Refs. \cite{Talbot,Privman} for the
review).

The problem of {\it competitive} RSA from a binary mixture is of
special interest because of the non-trivial correlations in both
the particle and gap-size distributions, developed during the
deposition. These correlations manifest themselves in the final
irreversible state corresponding to the so-called ``{\it jamming
limit}" --- when every gap capable of adsorbing a particle has
done so. Numerous studies, reported in the literature for the
binary-mixture RSA in the jamming limit, addressed the problem of
correlations only indirectly, through its manifestation in the
fill factor or the gap distribution. Available results include
binary mixtures with point-like particles \cite{bartelt,hassan1}
and those with a relatively small particle size ratio, $b/a<2$
\cite{hassan2}. Also available are Monte-Carlo studies of the fill
factor and the gap-size distribution for a binary-mixture
deposition with equal abundance of both particles \cite{Araujo}.

The present study is concerned with the correlation between the
fluctuations in the number of adsorbed particles of each kind from a
two-size binary mixture, as well as with their partial contributions
to the fill factor. We present both analytical results and those
obtained by Monte-Carlo simulations for a wide range of
binary-mixture compositions and size ratios.

We are interested in the RSA problem primarily because of its
relevance to the propagation of high-energy $\gamma$-particles
through a semiconductor crystal --- with particle energy branching
(PEB) due to cascade multiplication of secondary electrons and holes
\cite{ino,Devan,Spieler,Klein,rusb}. The correlation of energy
distribution between secondaries is quite similar to that of the gap
distribution in the RSA process \cite{corr}. In both cases, the
ratio of the variance of the final number of particles to the
average particle number in the final (jamming) state can be much
less than unity, which is favorable for the detector energy
resolution. This ratio (which would be unity if the particle number
obeyed a Poisson distribution) is called the Fano factor, $\Phi$
\cite{Fano}.

The reported attempts to evaluate $\Phi$ employed oversimplified
models of the semiconductor band structure. In such models, all
crystal properties are characterized by three parameters, namely,
the band gap, the phonon frequency, and the ratio of the rate of
phonon emission to that of impact ionization. The price of this
oversimplification had been that correspondence with experiment
could be achieved only by assuming unphysically large rates of
phonon losses (about $0.5$ eV per created e-h pair). This does not
corroborate with the known values for the ratio of the impact
ionization and the phonon emission probabilities for high-energy
electrons in semiconductors. The model furthermore obscures the role
of features in the band structure and the ionization process that
are specific to a particular semiconductor.

In our earlier work \cite{assl}, we used an extended RSA model of
particles that expand or shrink upon adsorption. The shrinking model
is relevant to the PEB problem in that it helps to elucidate such
factors as the non-constant density of states in the semiconductor
band and the fact that due to momentum conservation the ionization
threshold is larger than the actual (bandgap) energy that is lost in
impact ionization.

The recursive technique employed in Ref. \cite{assl} allowed us to
assess the accuracy of approximate approaches to the yield and
variance calculations (such as, e.g., the average-loss approach of
Refs. \cite{Spieler,Klein}).

In the present work, the RSA model is extended in a different
direction  --- competitive deposition of different-size particles
from a binary mixture --- that is suitable to simulate the role of
multiple channels of pair production, owing to the multi-valley
nature of semiconductor bands. We arrive at a number of qualitative
conclusions that should be taken into account in both the
interpretation of experimental data and the choice of the crystal
composition and device structure in gamma detectors optimized for
energy resolution.

The paper is organized as follows. Section II presents the basic
equations of the recursive approach and the analytical results for
the fill factor and its variance for the larger particles. In Sect.
III, we analyze the results that demonstrate high correlation in the
particle distribution. Based on the gained understanding, we
formulate in Sec. IV the implications of our results for the Fano
factor of semiconductor $\gamma$ detectors. Our conclusions are
summarized in Sect. V. Certain analytical results are derived in the
Appendix.

\section{Partial contributions to the fill factor and its variance for two-size RSA problem}
\label{fillFx} We consider the problem of competitive deposition
from a binary mixture of particles with sizes $a$ and $b$, whose
relative contributions to the total flux on the adsorbing line are
$q$ and $p=1-q$, respectively.  We shall use a recursive approach to
first study the mean number of particles $n_a(x)$ and $n_b(x)$,
adsorbed on a line of length $x$ (in the jamming limit), and then
the corresponding variances.

Consider a large enough empty length  $x > a,b$. We assume that
the adsorption is sequential, i.e. only one particle is adsorbed
at a time. The first adsorbed particle will be of size $a$ with
the probability of landing at any point  $q(x-a)/(x-\overline l)$
or of size $b$ with the landing probability $p(x-b)/(x-\overline
l)$.  Here $\overline l=qa+pb$ is the ``average" particle size in
the binary flux. After the first particle is adsorbed, it fills a
certain interval $[y,y+a]$ (or $[y,y+b]$), and leaves two
independent segments, whose combined size is either $x-a$ or
$x-b$. The average numbers of $a$-particles $n_a(y)$ and
$n_a(x-y-a)$ (or $n_a(y)$ and $n_a(x-y-b)$) will be subsequently
adsorbed in these gaps. Thus, the recursion relation is of the
form \ba n_a(x)&=& \frac {q(x-a)}{x-\overline l}
[1+n_a(y)+n_a(x-a-y)] \nonumber \\ &&+\frac {p(x-b)} {x-\overline
l} [n_a(y)+n_a(x-b-y)]~, \nonumber \ea  where the first and the
second terms (upper and lower lines) correspond to the cases of
the first landed particle being a particle of sort $a$ or $b$,
respectively. These cases must be averaged over all possible
landing coordinates $y$ of the first particle in a different way,
viz. for $a$ first,
$$<n_{a}(y)>_a=\frac{1}{x-a}\int_0^{x-a}n_{a}(y)dy~,$$
whereas for $b$ first,
$$<n_{a}(y)>_b=\frac{1}{x-b}\int_0^{x-b}n_{a}(y)dy ~.$$
Performing the average and using the symmetry between left and
right segments we obtain, finally:
 \ba n_a(x)&=&\frac {q(x-a)}{x-\overline
l}+\frac {2 q}{x-\overline l}\int_0^{x-a}n_a(y)dy \nonumber \\
&&+\frac {2 p}{x-\overline l}\int_0^{x-b}n_a(y)dy.  \label{1equ-a}
\ea A similar equation holds for the particles of size $b$: \ba
n_b(x)&=&\frac {p(x-b)}{x-\overline l}+\frac {2 q}{x-\overline
l}\int_0^{x-a}n_b(y)dy \nonumber \\ &&+\frac {2 p}{x-\overline
l}\int_0^{x-b}n_b(y)dy. \label{1equ-b} \ea With the help of Eqs.
(\ref{1equ-a},\ref{1equ-b}) one can readily derive an equation for
the average total covered length $f(x)$,  defined as
$f(x)=an_a(x)+bn_b(x)$, giving \ba f(x)&=&\frac {x\overline
l-qa^2-pb^2}{x-\overline l}+\frac {2 q}{x-\overline
l}\int_0^{x-a}f(y)dy \nonumber \\&& +\frac {2 p}{x-\overline
l}\int_0^{x-b}f(y)dy. \label{1equ-t}\ea Equation (\ref{1equ-t})
agrees with that of Ref. \cite{mao} for the total covered length in
RSA from a multi-size mixture. However, the advantage of Eqs.
(\ref{1equ-a},\ref{1equ-b}) is that they permit studying the partial
contributions to the coverage by each of the two sorts of particles
separately.

Note that the symmetry between the $a$- and the $b$- particles is
broken by the initial conditions. To be specific, let $b>a$. Then,
for $b$-particles the boundary condition at small $x$ is simply
 \be n_b(x)= 0, \
\hsize=1cm \  0\le x < b \label{in_b} \ee whereas for
$a$-particles we have
 \be n_a(x)=\left \{
\begin{array}{ccc} 0,&\hfill&0\le x < a \\ 1,&\hfill& a  < x \le
{\rm min}(2a ,b) \cr\end{array} \right.  \label{init_a} \ee For
$b>2a$, Eq. (\ref{init_a}) should be supplemented with  \be
n_a(x)=1+\frac{2}{x-a}\int_0^{x-a}n_a(y)dy \label{init_as} \ee Eq.
(\ref{init_as}) accounts for the deposition of smaller particles in
small gaps where the larger particle does not fit. Clearly, this
process is not influenced by the $b$-particles and does not involve
particle competition.

More refined  arguments are needed to derive the second moment of
the distribution, i.e. the expected value of the square of the
number of particles of a given sort, $u_a(x)={\rm E} n_a^2(x)$. It
may not be $a~priori$ evident that one can write independent
expressions for particles of both sorts, because parameters $a$
and $b$ not only describe the particle size but also designate the
sort of a particle. Indeed, we can even have $a=b$ and distinguish
the particles by some other parameter, like ``color''. Our
approach should remain valid in this case too. To be rigorous, we
therefore introduce an artificial parameter, the ``mass'' of a
particle, $m_a$ and $m_b$, whose value may depend on the particle
shape and is simply proportional to the particle length only for a
fixed transverse particle size. Hence one can regard $m_a$ and
$m_b$ as independent parameters.

Consider a total mass $M(x)=m_an_a(x)+m_bn_b(x)$ of the particles
adsorbed in a line segment $x$. We first evaluate recursively the
mean square of the total mass
$<M^2(x)>=\langle[m_an_a(x)+m_bn_b(x)]^2\rangle$, and then
calculate the second partial derivatives with respect to $m_a$ and
$m_b$. Using the landing probabilities of particles to perform the
averaging,  we obtain \ba u_a(x)&=&(x-\overline l)^{-1}\left[
q(x-a)+2 q\int_0^{x-a}u_a(y)dy\right.\nonumber
\\&&+2 p\int_0^{x-b}u_a(y)dy +4
q\int_0^{x-a}n_a(y)dy \nonumber \\&& +2
q\int_0^{x-a}n_a(y)n_a(x-y-a)dy \nonumber \\&& +\left.2
p\int_0^{x-b}n_a(y)n_a(x-y-b)dy\right]
 \label{1equ-va}\ea
Similarly, equation for $u_b(x)$ reads \ba u_b(x)&=&(x-\overline
l)^{-1}\left[p(x-b)+2 q\int_0^{x-a}u_b(y)dy\right. \nonumber
\\&&+2 p\int_0^{x-b}u_b(y)dy
 +4 p\int_0^{x-b}n_b(y)dy \nonumber
 \\&& +2 q\int_0^{x-a}n_b(y)n_b(x-y-a)dy
  \nonumber \\&&\left.+2 p\int_0^{x-b}n_b(y)n_b(x-y-b)dy\right]
 \label{1equ-vb}\ea

We could have derived Eqs. (\ref{1equ-a},\ref{1equ-b}) in a
similar way, by first evaluating the total average mass
$M(x)=m_an_a(x)+m_bn_b(x)$ recursively, and then calculating the
derivatives.
For a more general case, when the total mass is a linear functional
		$M(x)=\int m_l n_l(x) dl$ on the mass distribution $m_l$, one would
   	have to use variational derivatives $\delta M(x)/\delta m_l$.
 For the case of binary mixtures we consider, partial
derivatives are sufficient.

Similarly, we derive an equation for the correlation function
$u_c(x)=\langle n_a(x)n_b(x) \rangle$ by calculating a mixed
derivative of $<M^2(x)>$ with respect to $m_a$ and $m_b$. For
particles uniform in the transverse direction with unit mass
density, both the mass and the length of particles are identical,
which gives a way to check the equations. An appropriate linear
combination of equations for $u_a$, $u_b$, and $u_c$ then gives an
equation for the variance of the total filled length or,
equivalently, for the variance of the wasted length,
$w(x)=x-f(x)$. The resulting equation can also be obtained
directly, by applying recursion arguments to the waste. The
identical results obtained can be viewed as an additional proof of
Eqs. (\ref{1equ-va},\ref{1equ-vb}).

Note the asymmetry in the 4-th terms of Eqs.
(\ref{1equ-va},\ref{1equ-vb}) that are proportional, respectively,
to $4q$ and $4p$. These terms ensure the correct (linear) asymptotic
behavior of the variance at large $x$.

An important feature of Eqs.
(\ref{1equ-a},\ref{1equ-b},\ref{1equ-va},\ref{1equ-vb}) is that in
spite of the competitive character of the deposition of particles of
different sorts, the equations for  $n_a$, $n_b$ and the higher
moments are independent. This is rooted in the fact that a single
deposition step on an empty length $x$ does not depend on the
already adsorbed particle distribution.

Due to the self-averaging nature of the filling length (and waste
length) in the limit $x\rightarrow \infty$ the averaged (hence
approximate) recursion equations yield {\it exact} results. The
recursive technique is in this sense equivalent to the alternative
``kinetic" approach to RSA that is sometimes regarded as a
higher-level theory. In the kinetic approach one considers the rate
equation that describes the sequential deposition of particles with
the particle distribution on a line characterized by a
time-dependent function $G(x,t)$ representing the average density of
gaps whose size is between $x$ and $x+dx$ \cite{viot,hassan1}. It
has been ascertained for a number of problems that both approaches
give the same result for the coverage. Still, each has its own
benefits. The kinetic approach allows studying the temporal
variation of a state with specified particle distribution. The
recursive approach, while simulating a simplified version of the
kinetics, allows to study more complex effects, such as variance of
the adsorbed particles of different size.

\begin{figure} 
\epsfig{figure=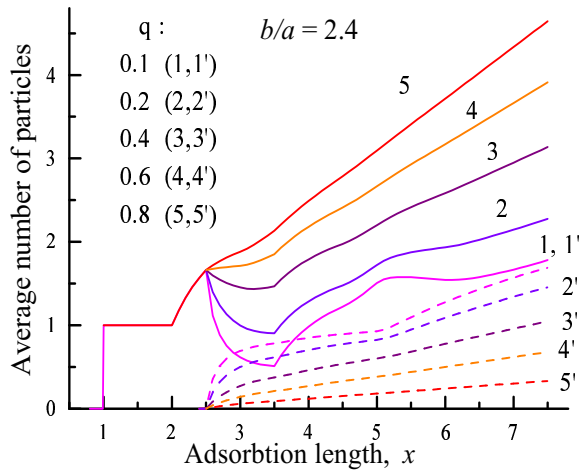,width=7.6cm,height=6.3cm}
\caption{(color online) Average number of adsorbed particles $n_a$
(solid lines) and $n_b$ (dashed lines) as functions of the length
$x$ (measured in units of $a$) of the adsorption interval, assumed
initially empty. The results are obtained by iterating Eqs.
(\ref{1equ-a},\ref{1equ-b}) with the assumed ratio of the particle
size $b/a=2.4$ and the varying fraction $q$ of $a$-particles in
the flux. } \label{3f}
\end{figure}

Evaluation of $n_a(x)$ and $n_b(x)$ is readily done by repeated
iterations of Eqs. (\ref{1equ-a},\ref{1equ-b}), going from the
small to progressively larger lengths $x$. Results of the
numerical recursion are shown in Fig. 1 for a particle size ratio
$b/a=2.4$ and varying $q$.

The noteworthy features of the functions $n_a(x)$ and $n_b(x)$ are
(i) the step-like features at $x=a$, $x=b$ (which are replicated
with ever smaller amplitudes at $x=na+mb$, where $n$ and $m$ are
integers), (ii) the dip in the number of small particles $n_a(x)$
at $x=b$, which increases with $p$, and (iii) the reduction of
$n_b$ with increasing $q$. We also note that for all $q$ the
behavior of both $n_a(x)$ and $n_b(x)$ becomes very close to
linear already at $x \approx 7$.

The asymptotic behavior of $n_a(x)$ and $n_b(x)$ at large $x$ can
be obtained by multiplying Eqs. (\ref{1equ-a},\ref{1equ-b}) by
$x-\overline l$ and taking the derivative with respect to $x$. The
resulting differential equations are satisfied by linear functions
of the form \be n_a=\alpha_a (x+\overline l)-q, \ \ n_b=\alpha_b
(x+\overline l)-p \label{asy_nab},\ee where $\alpha_a$ and
$\alpha_b$ are arbitrary constants. When correctly chosen (by
matching to the recursive solution) these constants become the
partial filling factors. After the matching is done, the total
filled length in the asymptotic limit is given by $f(x)=\theta x
+(\theta-1)\overline l, \label{fill_asy}$ where
$\theta=a\alpha_a+b\alpha_b$ is the specific coverage. It is
worthwhile to stress that the value of the asymptotic solutions
(\ref{asy_nab}) consists precisely in that they are asymptotically
exact. Hence they provide a sanity check on any solution we could
have obtained by a numerical recursion up to moderate values of
$x$.

Similarly, Eqs. (\ref{1equ-va}-\ref{asy_nab}) yield the variances
at large $x$,
\begin{subequations}
\label{asy_uab}
\ba u_a-n_a^2&=&\mu_a (x+\overline l)-qp[1+(b-a)\alpha_a]^2 ,\label{asy_ua}  \\
u_b-n_b^2&=&\mu_b (x+\overline l)-qp[1-(b-a)\alpha_b]^2
\label{asy_ub}.\ea \end{subequations} Again, these solutions are
asymptotically exact; they satisfy Eqs.
(\ref{1equ-va},\ref{1equ-vb}) with arbitrary values of $\mu_a$ and
$\mu_b$, provided of course that $n_a(x)$ and $n_b(x)$ are in the
correct asymptotic form (\ref{asy_nab}) with properly chosen [i.e.,
satisfying Eqs. (\ref{1equ-a},\ref{1equ-b})] coefficients $\alpha_a$
and $\alpha_b$. In principle, we could now follow a procedure
similar to above, viz. determine $\mu_a$ and $\mu_b$ by matching
Eqs. (\ref{asy_uab}) against a numerical recursive solution at some
moderate value of $x$. However, it would be rather difficult to
control the numerical accuracy in this procedure, because of the
difference of nonlinear functions that enter Eqs. (\ref{asy_uab}),
even though that difference itself behaves linearly with $x$ at
large $x$.

Fortunately, our model admits of an exact solution based on the use
of Laplace transformation (details can be found in \cite{assl} and
references therein).  Below we present an exact evaluation of
variance for particles of larger size, while details of similar
though lengthier calculations for smaller particles are presented in
Appendix.

Firstly, we need exact solutions of Eqs.
(\ref{1equ-a},\ref{1equ-b}). To obtain these, we substitute
$x\rightarrow x+b$ in Eq. (\ref{1equ-b}) and multiply it by
$x-\overline l$. Taking the Laplace transformation of the
resulting equation and using the boundary condition (\ref{in_b}),
we obtain \be \left[-\frac {d}{ds} +b-\overline
l\right]e^{bs}N_b(s)=\frac{p}{s^2}+\frac{2}{s}\left(q
e^{(b-a)s}+p\right)N_b(s) \label{fill_b} \ee Here $N_b(s)$ is the
Laplace transform of $n_b(x)$, \be N_b(s)=\int_0^\infty e^{-sx}
n_b(x)dx, \label{laplace} \ee Rearranging the terms and
multiplying by $e^{-bs}$, we put Eq. (\ref{fill_b}) into the form
\be N_b'(s)+\left[\overline
l+\frac{1}{s}\left(qe^{-as}+pe^{-bs}\right)\right]N_b(s)=
-\frac{p}{ s^2}e^{-bs}. \label{eq-ff} \ee For $p \rightarrow
\infty$, the solution of Eq. (\ref{eq-ff}) is, asymptotically, \be
N_b(s)|_{s \rightarrow \infty}=\frac{p }{s(b-\overline l)}e^{-bs},
\label{bound} \ee as follows from the known variation of
$n_b(x)\approx p(x-b)/(b-\overline l)$ at small $x-b$. Hence we
have \be N_b(s)=\frac{p \exp(-\overline
l~s)}{s^2\beta(s)}\int_s^\infty e^{-q(b-a)t}\beta(t)dt,
\label{eq-ff1} \ee where \be \beta(s)= \exp\left[- 2\int_0^{s}
\left(\frac{1- q\exp(-at)-p \exp(-bt)}{t}\right)dt\right].
\label{beta} \ee

To find the asymptotic behavior of $n_b(x)$ at large $x$, it is
convenient to use Karamata's Tauberian theorem for the asymptotic
growth rate of steadily growing functions (see e.g. \cite{Variat},
p. 37). According to the theorem, the asymptotics of $n_b(x)$ [or
$n_a(x)$ or their variances] can be readily obtained (by taking the
inverse Laplace transformation) from the Laurent power series
expansion of the Laplace transforms of these functions at small $s$
(see \cite{coff} for the mathematical details of this analysis).

Function $N_b(s)$ is analytic at all $s\ne 0$ and at $s=0$ it has a
second-order pole with the following asymptotic \be
N_b(s)=\frac{\alpha_{b,0} }{s^2}+\frac{\alpha_{b,0}\overline l
-p}{s}+O(s), \label{asympF} \ee where \be
\alpha_{b,0}=p\int_0^\infty e^{-q(b-a)s} \beta(s)ds. \label{filRen1}
\ee To calculate $n_b(x)$ at large $x$, we take the inverse Laplace
transformation of (\ref{asympF}).  This gives \be
n_b(x)=\alpha_{b,0}(x+\overline l)-p, \label{solu-n} \ee with an
exponentially small error term, in line with the asymptotics given
by Eq. ($\ref{asy_nab})$.

In the limit $p=1$, equation (\ref{filRen1}) duly gives the
so-called jamming filling factor $R$ for the standard RSA,
$\alpha_{b,0}(l=1) \equiv R=0.74759\cdots$ (also called the Renyi
constant \cite{renyi}). In the limit $a \rightarrow 0$, Eq.
(\ref{filRen1}) recovers the results of Refs.
\cite{bartelt,hassan1} for the coverage of a line from a binary
mixture of finite size particles and point defects. Moreover, Eq.
(\ref{filRen1}) gives the large particle contribution to the total
coverage, obtained in \cite{mao,hassan2} for the range $a< b< 2a$.
Here we see that this result remains valid for arbitrary $a<b$.

Next, we perform similar manipulations with Eq. (\ref{1equ-vb}) and
obtain an equation for the Laplace transform of the variance
$U_b(s)=\hat L [u_b(x)]$, viz. \be U_b'(s)+\left[\overline
l+\frac{2}{s}\left(qe^{-as}+pe^{-bs}\right)\right]U_b(s)=-
\frac{\exp(-bs)}{s^2}R_b(s), \label{eqM} \ee where \be
R_b(s)=p+4psN_b(s)+2s^2N_b^2(s)\left(qe^{(b-a)s}+p\right),
\label{Rb} \ee with $N_b(s)$ defined by Eq. (\ref{eq-ff1}). The
solution of Eq. (\ref{eqM}) can be written in a form similar to Eq.
(\ref{eq-ff1}), namely \be U_b(s)=\frac{\exp(-s \overline
l)}{s^2\beta(s)}\int_s^\infty \beta(t)e^{-q(b-a)t}R_b(t)dt.
\label{eq-ffM} \ee  The integrand in the right-hand side of Eq.
(\ref{eq-ffM}) is proportional to $1/t^2$ causing the integral to
diverge as $1/s$ for $s \rightarrow 0$. This is due to the
square-law dependence of $u(x)$ at large $x$ .

To separate the regular part needed for the estimation of variance,
we note that at small $t$ one has $R_b(t) \propto
\alpha_{b,0}t^{-2}$. Moreover, the series expansion shows that the
difference $\beta(t)\exp[-q(b-a)t]R_b(t)-2\alpha_{b,0}^2t^{-2}$ is
regular at $t\rightarrow 0$. Therefore, it is convenient to define
an entire function
$\kappa_b(t)=\beta(t)\exp[-q(b-a)t]R_b(t)-2\alpha_{b,0}^2t^{-2}$. In
terms of this function, the solution $U_b(s)$ can be expressed as
follows: \be U_b(s)=\frac{ \exp(-s\overline
l)}{s^2\beta(s)}\left[\frac{2\alpha^2_{f,0}}{s} +k_{b,0}-\int_0^s
\kappa_b(t)dt\right], \label{eq-ffren} \ee where \be
 \ \ k_{b,0}=\int_0^\infty \kappa_b(t)dt. \label{k-zero}\ee To apply
Karamata's Tauberian theorem, we note that the asymptotic expansion
of $U_b(s)$ near its third-order pole is of the form \ba
U_b(s)&=&\frac{2\alpha_{b,0}^2}{s^3}+\frac{k_{b,0}+2
\alpha_{b,0}^2\overline l}{s^2}\nonumber \\&&
+\frac{k_{b,0}\overline l-\kappa_b(0)-qp(b-a)^2\alpha_{b,0}^2}{s}.
\label{series-M} \ea Taking the inverse Laplace transformation, we
find the asymptotic form of $u_b(x)$: \ba u_b(x)& =& \alpha_{b,0}^2
x^2+ (k_{b,0}+2 \alpha_{b,0}^2\overline l) x\nonumber
\\&&+k_{b,0}\overline l-\kappa_b(0)-qp(b-a)^2 \alpha_{b,0}^2 ,
\label{FFF} \ea with an exponentially small error term. Using Eq.
(\ref{solu-n}) to subtract $n_b^2(x)$, we find an equation of the
form (\ref{asy_ub}) with $\mu_b=k_{b,0}+2p \alpha_{b,0}$. The
specific variance of the adsorbed number of $b$-particles is given
by (at $x\rightarrow \infty$)
 \ba \mu_b&=&\alpha_{b,0}(1+2p)+ 2\int_0^\infty
\left\{\beta(s)sN_b(s)e^{\overline l s}\left[2pe^{-bs}
\right.\right. \nonumber\\ &&\left.\left. +
sN_b(s)\left(qe^{-as}+pe^{-bs}\right)\right]
-\frac{\alpha_{b,0}^2}{s^2}\right\}ds. \ea Integrating by parts
the last term and rearranging the result, we finally obtain \ba
\mu_b&=\alpha_{b,0}(1-2p)+4p\int_0^\infty
\frac{\alpha_b(u)}{u}e^{-bu}\left(1-qe^{-au} \right. \nonumber \\
&\left. -pe^{-bu}\right)du +2\int_0^\infty
\frac{\alpha^2_b(u)}{\beta(u)u^2} e^{-\overline lu}K(u)du ,
\label{var_k} \ea where \ba
K(u)&=qe^{-au}\left[2(1-qe^{-au}-pe^{-bu})-(a+\overline l)u\right]
 \nonumber \\&+pe^{-bu}\left[2(1-qe^{-au}-pe^{-bu})-(b+\overline l)u\right]\ea and
\be \alpha_b(u)=\alpha_{b,0}-p\int_0^u e^{-q(b-a)y}\beta(y)dy
\label{alph_u}. \ee

In the limit of small $p\rightarrow 0$, the Fano factor $\Phi=
\mu_b/\alpha_{b,0}\rightarrow 1$. In this limit, large particles
are distributed on the line randomly, without correlations. In the
opposite limit, $p=1$, Eq. (\ref{var_k}) reduces to the standard
RSA result, first obtained for a lattice RSA model by Mackenzie
\cite{Mack}. The numerical value of the Mackenzie constant, $\mu_0
= 0.0381564\cdots$, corresponds to $\Phi =0.0510387\cdots$, see
\cite{coff}. Expression (\ref{var_k}) for the larger particles has
the same structure as the corresponding formula in the standard
RSA model (fixed-size CPP). Due to the exponential factors in the
integrands of Eq. (\ref{var_k}), the dependence of $\mu_b$ on $a$
for $a\ll b$ is quite weak. The limiting value of the specific
variance for $a/b \rightarrow 0$ gives the specific variance of
the fill factor for the case of finite-size particles ($b=1$)
mixed with point-size particles,
 \ba \mu_{b,p}&=\alpha_{b,p,0}(1-2p)+4p^2\int_0^\infty
\frac{\alpha_{b,p}(u)}{u}e^{-u}\left(1-e^{-u}\right)du \nonumber \\&
+2p\int_0^\infty \frac{\alpha^2_{b,p}(u)}{\beta_p(u)u^2} \left
\{qe^{-pu}\left(2-2e^{-u}-u\right) \right. \nonumber
\\&\left.
+e^{-(1+p)u}\left[2p\left(1-e^{-u}\right)-(1+p)u\right]\right\}du,
 \label{mu_p} \ea
where $\alpha_{b,p,0}$ is the fill factor for this case, \be
\alpha_{b,p}(u)=p\int_u^\infty e^{-qy}\beta_p(y)dy, \  \
\alpha_{b,p,0}=\alpha_{b,p}(u=0) \label{alpha_p}\ee and \be
\beta_p(u)=\exp\left[-2p\int_0^u
\left(\frac{1-\exp(-t)}{t}\right)dt\right]. \label{beta_p} \ee It
is worth to note that Eqs. (\ref{1equ-b},\ref{1equ-vb}) and their
solutions can be readily generalized to the case when particles of
the smaller size have an arbitrary distribution in the interval
$[a_1,a_2]$ so long as $a_2\le b$ \cite{asslr}.

The above analytic results for the variance of larger particles are
essentially exact, as will be confirmed in the next Section by Monte
Carlo simulations. For the smaller particles, the calculations are
messier and accurate analytical results can be obtained only in a
certain range of particle size ratios. Estimations of the variance
for small-size particles are further discussed in the Appendix.

\section{Discussion of the results, comparison with Monte Carlo modeling}
Here we present the results of numerical calculations using both the
analytical expressions obtained in the preceding section and Monte
Carlo simulations. For large-size particles the Monte Carlo results
are very close to analytical expressions both for the fill factor
and the variance, so we shall not dwell on their comparison. For
small-size particles, especially in the range $2<b/a<8$, analytical
calculations are rather unwieldy, so Monte Carlo simulations become
indispensable. Larger size ratios, $b/a>8$, lend themselves to an
approximate analytical approach (see Appendix). In this case, we use
the Monte Carlo to estimate its accuracy for the small particle
contribution.

Traditional studies of the generalized RSA via Monte Carlo
simulations follow a temporal sequence of events. For the case of
adsorption on a line of the length $x$ from a binary mixture, one
step of the sequence comprises:

(i) selection of a particle from the mixture according to the
deposition flux ratio (with the probability $q$ of choosing the
small-size ($a$) particle, and the probability $p=1-q$ of selecting
a particle of larger size $b$);

(ii) random choice of a deposition coordinate of particle center on
the line $x$ with formerly deposited particles;

(iii) rejection of the particle if it overlaps by any part with
formerly deposited particles or with the line borders; otherwise,
the particle deposition proceeds with the formation of two new
disconnected adsorption lengths.

This traditional approach has several drawbacks, that make the
modeling very demanding, both in terms of the computer time and
memory allocation.

Firstly, both the filled length in the jamming limit and the
specific fill factor (coverage) depend on the initial length. Due
to the self-averaging property of the coverage it tends to a
unique exact value in the limit $x \rightarrow \infty$.  To obtain
the accuracy of about $0.1$ \%, the common strategy has been to
use large initial length values (10$^5b$ -$10^7b$) and make
additional averaging over a set of about $N_R=$100-1000 different
realizations.

Secondly, as time evolves and the jamming limit is approached, the
probability of finding a free gap for particle deposition becomes
greatly reduced, so that the adsorption time tends to infinity. The
process is terminated when variations of the adsorbed particle
number are smaller than those required by the desired accuracy.

The recursive analysis of the generalized RSA suggests a revision of
the above scheme. Since the deposition is random and sequential, it
does not depend on the temporal history of the process or the
growing number of rejected particles and their coordinates.
Therefore one step of the sequence can be chosen as follows:

(i) selection of any free deposition length, $l_1 >a$. It is
convenient to choose for $l_1$ the outermost free deposition
length on the left-hand side.

(ii) if $l_1<b$, then particle of size $a$ is deposited, otherwise
the deposited particle is chosen according to the landing
probability, given by $q(l_1-a)/(l_1-\overline l)$ for
$a$-particle and $p(l_1-b)/(l_1-\overline l)$ for $b$-particle,
where
 $\overline l=qa+pb$.

(iii) random choice of a deposition coordinate (taken as the
coordinate of particle's left end) on the line $l_1$ for a given
particle size, i.e. within the interval $l_1-a$ for $a$-size or
within $l_1-b$ for $b$-size particle, with the formation of two new
adsorption lengths from the initial length $l_1$.

It is readily seen that although the sequence of deposition events
is different from the actual temporal sequence of adsorption (the
simulated deposition proceeds by sequentially filling the left-hand
lengths), the statistics of divisions is identical and therefore so
is the final distribution of the gaps, as well as all statistical
properties of the jamming state. Our sequential scheme excludes
deposition of to-be-rejected particles and therefore is incomparably
faster. Besides, it terminates exactly when the jamming limit (with
no gaps larger than unity) is achieved. Direct comparison with the
traditional Monte Carlo results, e.g. \cite{hassan1,Araujo,mao}
exhibits total agreement. The difference in the calculation time is
especially evident for small (close to zero) $q$: in the time scale
of ``real" deposition, the jamming limit will be strongly delayed
because of the rarity of events with small particle chosen. In our
modified approach, all gaps smaller than $b$ are ``rapidly"
populated by small-size particles, however small be the value of
$q$.

The next step of the revision is to exploit the fact (proven
analytically in the preceding section) that in the jamming limit,
the linear dependence on the adsorption length of both the average
filled length and its variance is exponentially accurate, starting
from a reasonably short length, certainly not exceeding $x \approx
10 b$. Since this linear dependence has only two parameters
[actually only one, as the parameter ratio is exactly fixed by
analytical considerations, Eqs. (\ref{asy_nab},\ref{asy_uab})],
both the coverage and the variance can be determined with Monte
Carlo simulations of short samples.

To be sure, in order to achieve the same accuracy as that obtained
for long samples, the results should be averaged over a sufficient
number of realizations $N_R$. This, however, takes little memory
or time. Calculations show similar accuracy for different $x$ and
$N_R$, so long as their product $x\times N_R$ is fixed. The
results presented below were obtained using a sample of size
$x=200a$ for $b/a<10$ and $x=400a$ for $b/a=20,40$, subsequently
averaged over 10 000 realizations, which appeared to be sufficient
to eliminate any spread of the results in the graphical
presentation (producing an accuracy of better than 0.1\%).

The use of small samples is very effective in reducing the
calculation time (with an ordinary PC, high-accuracy results can be
obtained in minutes, compared to days in the traditional scheme
\cite{mao}).

\begin{figure} 
\centerline{\epsfig{figure=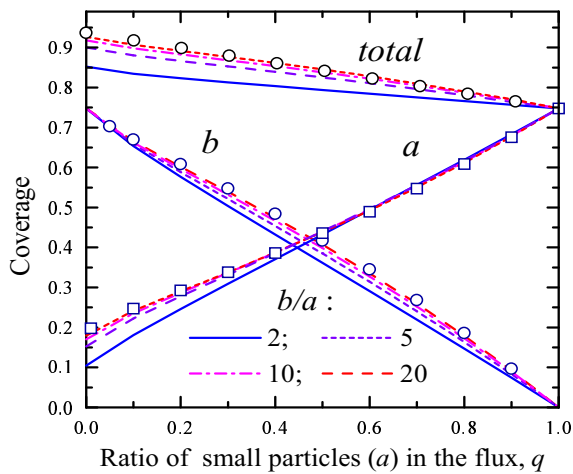,width=7.6cm,height=6.3cm} }
\caption{(color online) Partial contributions of small and large
particles to the total coverage depicted as functions of $q$, for
different particle-size ratios $b/a$ in the flow. Open points
correspond to the limit $b/a \gg 1$, as described by the
analytical formulae (\ref{alpha_p}) for $b$-particles and
(\ref{poin-smco}) for $a$-particles. For the total coverage, the
open circles to the wasted length product approximation, Eq.
(\ref{thetaRq}).} \label{3f}
\end{figure}

Figure 2 shows partial contributions to the coverage as functions of
the fraction $q$ of small particles in the binary mixture at
different ratios of particle size. As $q$ increases, the coverage
with large particles is substituted by that with small particles,
producing some decrease in the total coverage. In the regions of
corresponding parameters, our results reproduce those of reported
analytical calculations (i.e. for $b/a<2$ \cite{hassan1,mao} and for
$a=0$ \cite{hassan2} for the large particle contribution) and those
obtained by the Monte Carlo simulations of
\cite{hassan1,Araujo,mao}, demonstrating the validity of our revised
approach.

\begin{figure} 
\centerline{\epsfig{figure=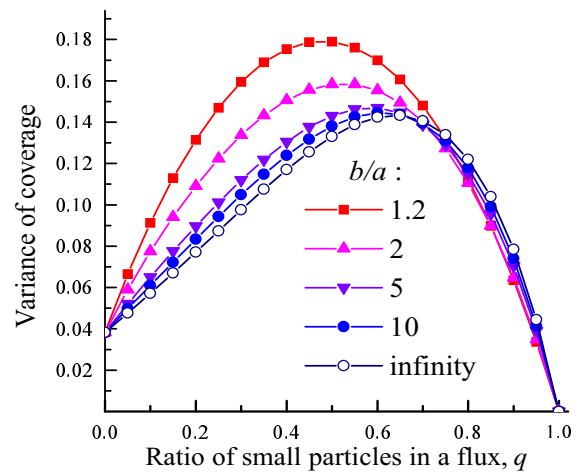,width=7.6cm,height=6.3cm} }
\caption{(color online) Variance of the partial coverage by
adsorbed $b$-particles from the binary mixture for different
values of the particle size ratio $b/a$ in the flow. } \label{3f}
\end{figure}

It is evident from Fig. 2, that the total coverage increases at
smaller $q$, as can be explained by sequential deposition of the
two kinds of particles. In the regime of small $q$, large
particles are adsorbed first and their deposition, unobstructed by
small particles, is tight. Subsequently, the small particle fill
the gaps between large particles and this clearly reduces the
total wasted length.

The effect of increasing the particle size ratio $b/a$ is
pronounced only for $b/a<10$, then it rapidly saturates.
Therefore, for large $b/a$, say $b/a=20$ the coverage by large
particles is very close to that obtained for a model mixture of
point-like and finite-size particles [by formally letting $a=0$ in
Eq. (\ref{filRen1})]. Such a model, however, has little relevance
to any practical situation, because it simply ignores the partial
contribution of small particles to the total coverage. The latter
can be described analytically in the limiting case $b/a
\rightarrow \infty$, Eq. (\ref{poin-smco}).

The partial contribution of small particles steadily grows with
the increasing size ratio due to the expanding gaps between the
large particles. In the limit $q \rightarrow 0$, the total
coverage can be estimated by observing that the specific wasted
length in this case is a simple product of the specific lengths
wasted in initial deposition of large particles and subsequent
deposition of small particles, i.e.
$1-\theta=(1-\theta_a)(1-\theta_b)$. Since for $q=0$ the specific
coverage $\theta_b=R$ and since for large size ratios (when the
gaps between large particles are large) the specific coverage
$\theta_a=R$, we have $\theta=1-(1-R)^2$=0.936, in agreement with the
results reported in the literature \cite{hassan1,Araujo}. However,
the sequential nature of the deposition suggests that the entire
$q$ dependence of the total $\theta$ can also be approximated by a
product of the specific wasted lengths in the competitive
deposition of large particles $q(1-R)$ and subsequent deposition
of small particles in the remaining gaps, which gives \be
\theta=1-(1-R)(1-pR)=R[1+p(1-R)]. \label{thetaRq} \ee This
product-waste approximation is shown in Fig. 2 by the open
circles.

Next, we concentrate on the specific waste variance and the Fano
factor. We shall discuss the $b$- and $a$-particles separately,
since the effects are rather different in nature and also since they
have been evaluated by different techniques. Results for large
particles are obtained by numerical integration of Eq.
(\ref{alph_u}) and confirmed by Monte Carlo simulations. Results for
$a$ particles are obtained by Monte Carlo stimulations and are
accompanied by analytical expressions in the limit $b/a\gg 1$.

Variance, $\tilde \mu$, of the partial contribution of
$b$-particles to the total coverage is shown in Fig. 3 for
different particle size ratios. Unlike the particle number
variance $\mu$, the variance of coverage, $\tilde \mu =\mu b$,
depends only on the size ratio $b/a$ and does not directly scale
with $b$. It is therefore more indicative of the effect of
decreasing size of small particles on the fluctuations of the
number of large particles. At $q\rightarrow 0$, when the
adsorption of large particles is unconstrained by small particles,
the variance of large particles is minimal and corresponds to the
highly correlated distribution \cite{corr} in the standard CPP
problem (one-size RSA). The variance rapidly increases with $q$ as
the small particle deposition destroys the CPP correlations. The
maximum of this effect is shifted to larger $q$ values for larger
$b/a$. For $q$ approaching unity, the variance decreases simply
due to the decrease of the average number of adsorbed
$b$-particles.

\begin{figure} 
\centerline{\epsfig{figure=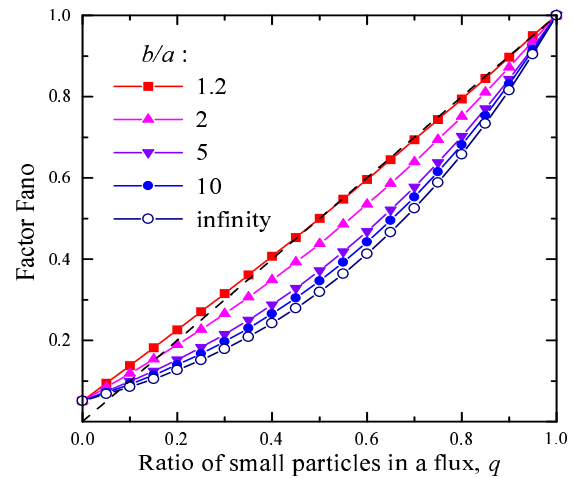,width=7.6cm,height=6.3cm} }
\caption{(color online) The Fano factor for the number of adsorbed
$b$-particles from the binary mixture as functions of $q$, for
different values of the ratio $b/a$ of particle sizes in the flow.
Dotted line corresponds to a random particle packing on a lattice
with suitable lattice constant ({\it aka} monomer adsorption). }
\label{4f}
\end{figure}
Correlation effects are more adequately characterized by the Fano
factor $\Phi_b$, shown in Fig. 4. With the increasing number of
competing small particles in the flux, the Fano factor grows from
the smallest value $\Phi =0.051\cdots$, corresponding to the
one-size RSA problem, to unity in the limit $q\rightarrow 1$.
Small coverage by the large particles in the latter limit means
that they are distributed randomly on the line, so that Poisson
statistics recovers. The most noticeable effect is a rapid
decrease of the Fano factor with $1-q$, manifesting a strong
enhancement of the correlation effects in the large particle
distribution. These correlation effects become exhausted only near
$q \le 0.1$. The correlation effects increase with $b/a$ but
saturate at about $b/a=20$.

\begin{figure} 
\centerline{\epsfig{figure=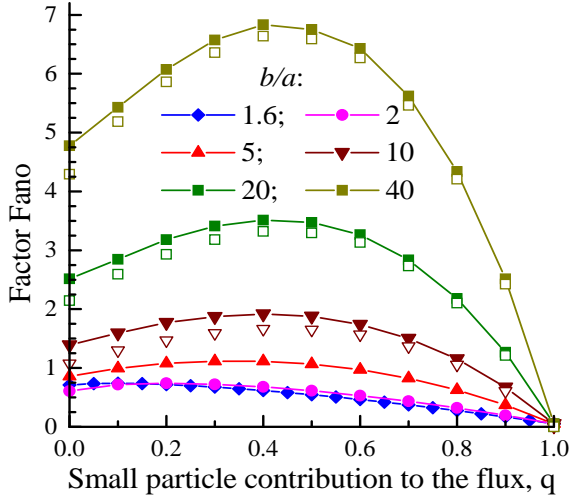,width=7.6cm,height=6.8cm} }
\caption{(color online) The Fano factor for adsorbed $a$-particles
from the binary mixture for different values of the particle size
ratio $b/a$ in the flow. Open points show the contribution of
fluctuations of the gap sizes between large particles.} \label{5f}
\end{figure}

Figure 5 shows the Fano factor for $a$-particles competitively
deposited along with large particles. The results are strikingly
different at all $q \ne 1$ (when $\Phi_a=\Phi$, as expected).
While the distribution remains correlated ($\Phi_a \le 1$) for
small ratios $b/a\le 5$, at larger $b/a$ one has  $\Phi_a
> 1$, almost for all $q$, which means that the number of small
particles per unit length is strongly fluctuating. This is due to
the widely fluctuating size of the gaps available for small
particle deposition between large particles. For large values of
$b/a$ and in the entire range of $q$, the Fano factor $\Phi_a$ can
be approximated in terms of the fluctuations of the coverage by
the large particles, viz. $\Phi_a=(b/a)\mu_{b,p}R^2/\theta_a$,
where $\mu_{b,p}$ is given by Eq. (\ref{mu_p}) and $\theta_a$ by
Eq. (\ref{poin-smco}). This approximation, which neglects
fluctuations of the density of adsorbed $a$ particles in the gaps,
is shown in Fig. 5 by open points. This contribution is
proportional to $b/a$ and for $b/a>10$ it is evidently dominant.

For the particle energy branching process at small $b/a<2$, both
the variance of the partial numbers of small and large particles
and the total number variance are of importance. We shall
illustrate this point in the instance of $b/a=1.2$ shown in Fig.
6. We see that at $q \approx 0.5$ the fill factor fluctuations are
larger for $a$ particles and somewhat smaller for $b$ particles,
but both are pretty large, compared to the variance of the total
number of adsorbed particles. This is due to the strong
anti-correlation in their distribution, as evidenced by the
specific fluctuation correlation function, $f_{cor}=x^{-1} \langle
\delta n_a \delta n_b \rangle$, also plotted in Fig. 6. We note
that $f_{cor}<0$, which means that any excess in the number of
$a$-particles is accompanied by a downward fluctuation in the
number of adsorbed $b$-particles. Importantly, the variance and
the Fano factor for the total number of adsorbed particles does
not exceed substantially its value for the single-size RSA.

\begin{figure} 
\centerline{\epsfig{figure=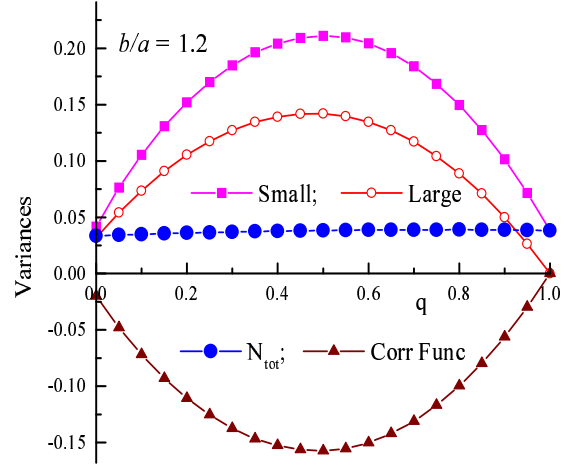,width=7.6cm,height=6.3cm}}
\caption{(color online) Variance of the partial number of adsorbed
$a$- and $b$-particles and of the total number of adsorbed
particles for $b/a=1.2$. Also shown is the fluctuation correlation
function $f_{cor}$.} \label{6f}
\end{figure}

Note the asymmetry of the curves for $a$ and $b$ particles, e.g.
the variance of large particles goes to zero as $q \rightarrow 1$
whereas that of small particles remains finite even as $q
\rightarrow 0$. This is a feature of our model that allows
"infinite" amount of time for the deposition of small particles in
the gaps left after the deposition of large particles is
completed, but not vice versa. Therefore, the deposition of small
particles remains finite even in the limit of $q \rightarrow 0$
and the same is true for the $a$-particle number variance.

Another interesting feature of the $a$-particle number variance,
already evident from Fig. 5, is its non-monotonic behavior as
function of $b/a$ at small $q$. This variation is displayed
directly in Fig. 7 that shows the dependence of the Fano factor on
$b/a$ for $q$=0.05, 0.1 and 0.2 --- where its non-monotonic nature
is most pronounced. The minimum of the Fano factor is achieved at
$b/a\approx 2$. Note that the non-monotonic dependence of the Fano
factor is accompanied by non-monotonic variations in the
dispersion of the gaps between small particles. In Ref.
\cite{Araujo} it was found that for $q=$0.5 the dispersion is
noticeably reduced at $b/a\approx 1.55$. These effects were
interpreted as a manifestation of the so-called ``{\it snug fit}"
events, i.e. particle deposition in gaps that are just barely
above the unit length $a$. In contrast, the Fano factor for
$b$-particles and that for the total number of particles remain
monotonic everywhere.

\begin{figure} 
\centerline{\epsfig{figure=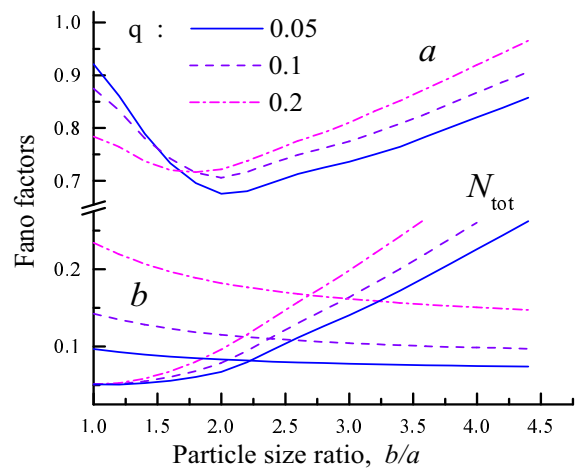,width=7.6cm,height=6.3cm}}
\caption{(color online) The Fano factor for adsorbed $a$- and
$b$-particles as functions of the particle size ratio $b/a$. Also
shown is the Fano factor for the total number of adsorbed
particles.} \label{6f}
\end{figure}

\section{Some consequences for the energy branching in high-energy particle detectors}
The model of RSA from binary mixtures is relevant to an important
practical problem of particle energy branching (PEB) where
high-energy particle propagates in an absorbing medium and
multiplies producing secondary electron-hole (e-h) pairs.
Multiplication proceeds so long as the particle energy is above the
impact ionization threshold \cite{Spieler}. The energy distribution
of secondary particles is random to a good approximation.

The affinity between the two problems was fully recognized already
in 1965 by van Roosbroek \cite{rusb} (see also \cite{Alkhas}). The
PEB process can be considered in terms of a CPP if one identifies
the initial particle kinetic energy with an available parking length
and the pair creation energy with the car size. Similarly, the
kinetic energies of secondary particles can be identified with the
new gaps created after deposition of a particle. Full equivalence of
PEB to CPP further requires that only one of the secondary particles
takes on a significant energy, which corresponds to binary cascades
\cite{Nay}. Otherwise, one has to consider a simultaneous random
parking of two cars in one event.

To estimate the particle initial energy in PEB, one measures the
number $N$ of created electron-hole pairs. Variance of this number,
due to the random character of energy branching and also due to
random energy losses in phonon emission, limits the accuracy of
energy measurements. Both the yield $\overline{N}$ and the e-h pair
variance ${\rm var} (N)=\overline{(N-\overline{N})^2}$ are
proportional to the initial energy. The ratio of the e-h pair
variance to the yield, i.e. the Fano factor of the PEB process, is a
parameter that quantifies the energy resolution of high-energy
particle detectors.

For semiconductor crystals, the PEB problem has additional
complications due to the energy dependence of phonon losses and the
energy dependence of the electron density of states and the impact
ionization matrix element. Full quantitative analysis of the PEB is
possible only with detailed numerical calculations, which goes far
beyond the scope of the present article.

A common feature of the energy branching process in semiconductors
is the presence of several pair production channels, associated
with the multi-valley energy band structure of the crystal. In Si,
Ge and common A$_3$B$_5$ semiconductors, the e-h pair creation
produces electrons in one of the ellipsoids near the edge of the
Brillouin zone, in 100 (X) or 111 (L) directions. Owing to the
difference in the final densities of states and the matrix
elements, the impact ionization processes associated with X and L
valleys have different but competitive probabilities. Because of
its low density of states, the $\Gamma$ valley is usually not
competitive, even when it is the lowest valley.

Ultimately, electrons will end up in the lowest energy valley but
when the final electron valley is itself degenerate, as in Ge or Si,
the resulting electron states may not be fully equivalent, because
of the different collection kinetics owing to the crystal
anisotropy. This effect may have important consequences for the
observed variance. For example, in Si diode detectors electrons are
created in 6 degenerate energy valleys that represent ellipsoids of
revolution elongated along (100) and equivalent directions in
k-space. Suppose the diode structure is such that the current flows
along the (100) direction, as it is usually the case. Electrons from
the two valleys along the current have a large mass and low
mobility. The measured current is hence dominated by electrons from
the 4 valleys elongated perpendicular to the current that have a low
mass and high mobility along the current. Since the choice of
equivalent valley in the PEB process is fully random, the number of
high-mobility electrons will fluctuate more strongly than the total
number of generated carriers. These fluctuations will dominate if
the inter-valley transition rate is low compared to the inverse
collection time. In the opposite limit of high inter-valley
transition rates, this effect will average out as the collected
current will fluctuate in time. The current fluctuation mechanism
due to the carrier escape into heavy-mass valleys is a well-known
source of noise in multi-valley semiconductors \cite{Kogan}). More
detailed account for these effects will be presented elsewhere
\cite{asslr}.

Here we shall discuss an opposite situation, that is common to {\it
direct-gap} semiconductors, such as GaAs or InP. In these materials,
the lowest ($\Gamma$) electron valley has a very low density of
states, compared to that in the satellite (X and L) valleys.
Therefore, the probability of electron generation in the
$\Gamma$-valley can be neglected in first approximation, so that the
branching competition occurs only between the satellite valleys of
two different kinds. Both the density of states and the threshold
energy are different between X and L valleys and we can use the
results of the present study to interpret and predict the
consequences, at least qualitatively.

The binary-mixture RSA model interprets the higher density of states
as higher deposition rate and the higher threshold as larger
particle size. To make our conclusions more transparent, let us
re-formulate the required results in terms of a random parking
problem with cars of two sizes. We are now interested only in the
numbers of parked cars and the fluctuations of these numbers.

Several qualitative conclusions can be drawn from our results:

(i) The total number of parked cars (in the jamming state) will
decrease with increasing fraction of larger cars in the flow and
with the growth of their size. For $b/a=1.4$ the effect is
illustrated in Fig. 8 (which can be viewed as an extension of Fig.
2). It follows from the fact that adsorption of a large car
excludes larger length for subsequent parking events and thus
causes a decrease of the total fill factor. Note that the decrease
in the total particle number is accompanied by an increase in the
total filled length, as smaller number of cars cover larger area.

\begin{figure} 
\centerline{\epsfig{figure=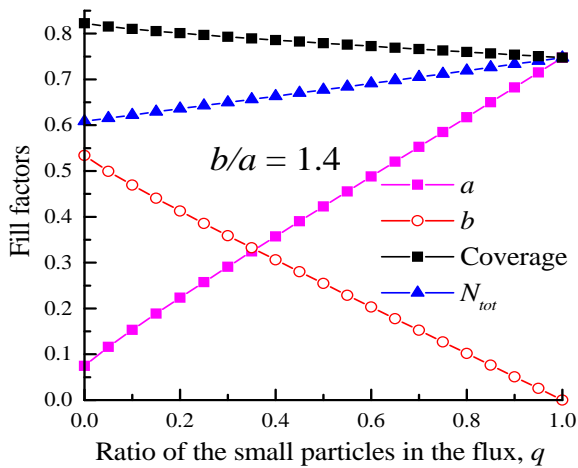,width=7.6cm,height=6.3cm}}
\caption{(color online) Partial fill factors and the total
coverage for $b/a=1.4$ as a function of $q$. Also shown is the
total number $N_{tot}$ of adsorbed particles} \label{6f}
\end{figure}

The next two conclusions (ii) and (iii), illustrated in Fig. 6, are
interconnected and will be discussed jointly.

(ii) Variance of the total number of parked cars and the Fano factor
will both grow with the increasing fraction of larger cars in the
flow and with the growth of their size.

(iii) Variance of the {\it separate} numbers of parked  small and
large cars and their Fano factors are considerably  larger than that
of the total number of cars. Therefore, if for some reason one type
of cars is neglected or undercounted, the registered variance and
the Fano factor can be substantially increased.

These conclusions are connected with the nature of the car number
fluctuations and the strong anti-correlation between the
fluctuations in the number of small and large cars. Fluctuations in
the number of parked cars of one kind are strongly enhanced by the
presence of more or less randomly distributed cars of the second
kind, especially when cars of the second kind dominate. This leads
to conclusion (iii). However, the two distributions are
anti-correlated (higher number of parked small cars is accompanied
by a smaller number of large cars and {\em vice versa}). The
anti-correlation is particularly strong for a size ratio that is
close to unity.

One can imagine a case when the two kinds of cars differ only in
``color".  In this case, Eqs. (\ref{filRen1},\ref{var_k}) yield
$\alpha_{a,0}=qR$, $\alpha_{b,0}=pR$, $\mu_a=Rqp+q^2\mu_0$ and
$\mu_b=Rqp+p^2\mu_0$, so that at large $x$ we have $<\delta n_a
\delta n_b>/x =-(R-\mu_0)qp$. Then, the anti-correlation is almost
complete: the fluctuations of the total number are much smaller
than those of a given color, but still non-zero. Both the
individual-color number fluctuations and the anti-correlation are
largest at $q\approx 0.5$, cf. Fig. 6. The anti-correlation
decreases with increasing size ratio, as reflected in our
conclusion (ii).

To discuss the above conclusions in terms of the PEB problem, we
note that estimation of the initial particle energy is equivalent
in CPP to a measurement of the unknown length of a parking lot in
terms of the total number of cars that were able to fit into it by
random parking, assuming that the average fill factor for a given
two-size car mixture is known from earlier measurements. The
absolute accuracy of such a measurement depends on the variance of
the fill factor, and the relative accuracy is determined by the
Fano factor. As shown above for a mixture of cars, the larger
disparity of car sizes leads to the higher fill-factor variance
and therefore reduces the absolute accuracy.

A particle detector measures the total number of secondary
particles of all sorts (but not their total creation energy, that
would be equivalent to the filled length). In any channel, all
secondaries that have sufficient energy for further branching will
do so. Therefore, only those pair creation energy ratios that
leave the channels competitive (i.e. $b/a<2$) are relevant to the
PEB problem
--- otherwise additional energy branching would be possible.

We conclude that the presence of competing channels with different
energies [e.g. impact ionization with excitation in X and L
valleys] will decrease the quantum yield (the number of
secondaries per unit energy of the primary particle) and enlarge
the Fano factor. The attendant loss in energy resolution is not
that bad when the ionization energies associated with different
valleys are not too disparate. For example, in Ge besides the
lowest eight L valleys ($E_G=0.66\:$eV) one has a non-competitive
$\Gamma$ valley ($E_{\Gamma}=0.8\:$eV) and six very competitive
Si-like valleys ($E_G=0.85\:$eV). The downgrading of energy
resolution should be more important for crystals with larger
($\approx 2$) threshold energy ratio. For example, in Si one has
besides the 6 lowest valleys ($E_G=1.12\:$eV) in X direction,
eight germanium-like L valleys with the gap $E_L=2.0\:$eV. Their
effect on the Fano factor in silicon may not be negligible.

Finally, reformulating  (iii), we stress that any significant
disparity in the collection efficiency between different
equivalent valleys will strongly enhance the Fano factor and
downgrade energy resolution. This happens because any collection
disparity breaks the symmetry between the equivalent valleys and
destroys the anti-correlation, responsible for keeping the total
Fano factor low even when the partial particle numbers associated
with individual valleys exhibit fully random fluctuations. One
possible origin for the asymmetry in the collection efficiency in
semiconductors has been discussed above in the case of silicon
diodes with the electric field in (100) direction. In germanium
diodes all different valleys are equivalent relative to the (100)
direction and the symmetry is not broken. It would be broken,
however, if one were to use Ge diodes oriented in (111) direction.
This would lead to a situation similar to Si --- with a possible
degradation in the Fano factor. These effects deserve additional
study, both experimental and theoretical.

\section{conclusions}
We have studied a generalized 1-dimensional competitive random
sequential adsorption problem from a binary mixture of particles
with varying size ratio. Using a recursive approach, we obtained
independent equations for the number of adsorbed particles of given
sort and exact analytical expressions for the partial filling
factors and variances for the larger particles. For the smaller
particles analytical expressions were obtained in a number of
limiting cases. The results have been confirmed by direct Monte
Carlo simulations. To do so, we have introduced a modified Monte
Carlo procedure that enabled us to explore a wide range of particle
size ratios and particle fractions in the flux.

A number of qualitative implications have been formulated, relevant
to the energy branching problem in high-energy particle propagation
through a semiconductor crystal. Conclusions made concern the
quantum yield and the energy resolution in semiconductor detectors
made of crystals with several competing channels of impact
ionization with different final electronic states.

We have found a very strong anti-correlation effects which strongly
suppress fluctuations of the total particle number compared to the
fluctuations of partial contributions by particles of a given sort.
This effect is particularly evident when one considers the
deposition of similar competing particles, e.g. parking of cars that
are different only in ``color". It may have dramatic consequences
for semiconductor $\gamma$-radiation detectors, if the symmetry
between anti-correlated particles is broken by a biased collection.
This leads to an important conclusion that the energy resolution of
semiconductor detectors is very sensitive to the collection
efficiency of competing secondary particles.

We have also found a very strong correlation effects that suppress
fluctuations of the larger particle number for all particle
ratios. As a result, the Fano factor for the larger particles is
as a rule considerably smaller than that for the smaller
particles. The variance of the coverage by the smaller particles
strongly increases with the growth of the particle size ratio
$b/a$. This effect is due to the fluctuations in the size of gaps
between larger particles that serve as receptacles for
small-particle deposition. For $b/a \ge 5$ the small-particle
variance exceeds that for the Poisson distribution in almost the
entire range of particle fractions in the flux onto the adsorbing
line.

 {\bf Acknowledgement.} This work was supported
by the New York State Office of Science, Technology and Academic
Research (NYSTAR) through the Center for Advanced Sensor
Technology (Sensor CAT) at Stony Brook.

\appendix
\section{Small particle contributions to coverage and coverage variance}
To calculate the contribution of small particles to the total
coverage at large $x$, we use Eq. (\ref{1equ-a}) with the initial
boundary conditions (\ref{init_a}). With the substitution
$x\rightarrow x+b$ and using Eq. (\ref{init_as}), we rewrite Eq.
(\ref{1equ-a}) in the form \ba (x+b-\overline
l)n_a(x+b)=q(b-a)n_a(b)+q x\nonumber
 \\+2q\int_{b-a}^{x+b-a}n_a(y)dy +2p\int_0^x n_a(y)dy
\label{Alpp-1}\ea Equation (\ref{Alpp-1}) is valid for all $x\ge b$.
Taking Laplace transformation of $n_a(x)$ cut at $x<b$ by a
step-function factor, we find that the transform, \be \tilde N_a
(s)=\int_b^\infty e^{-sx} n_a(x)dx, \label{laplaceNa} \ee satisfies
the following equation \ba \left(-\frac{d}{ds}+b-\overline
l\right)e^{bs}\tilde N_a(s)=\frac {q}{s^2}[1+(b-a)n_a(b)s]\nonumber
\\+2\frac{q}{s}e^{(b-a)s}\left[\tilde
 N_a(s)+J_1(s)\right]+2\frac{p}{s}\left[\tilde N_a(s)+J_2(s)\right]
\label{Lap-n1}\ea Here \be J_1(s)=\int_{b-a}^b e^{-sx} n_a(x)dx, \
\ J_2(s)=\int_{0}^b e^{-sx} n_a(x)dx \label{Jn1-Jn2}\ee
Rearranging the terms, we rewrite it in form \be
\left[\frac{d}{ds}+\overline
l+\frac{2}{s}\left(qe^{-as}+pe^{-bs}\right)\right]\tilde
N_a(s)=-\frac {1}{s^2}e^{-bs}R_a(s) \label{Lap-n2}\ee where \be
R_a(s)=q[1+(b-a)n_a(b)s]+2s\left[qe^{(b-a)s}J_1(s)
+pJ_2(s)\right]\label{R a}\ee The form of  Eq. (\ref{Lap-n2}) is
similar to Eq. (\ref{eqM}) in which, however $R_a$ should be
calculated through $J_1(s)$ and $J_2(s)$, using Eqs.
(\ref{init_a},\ref{init_as}). For the case $b<2a$ we have
$n_a(b)=1$, and $J_1(s)=J_2(s)$, while the explicit  expression
for $J_1(s)$ is easily obtained by substituting $n_a(x)=1$ in Eq.
(\ref{Jn1-Jn2}). Solution of Eq. (\ref{Lap-n2}) then enables one
to retrieve the result of Ref \cite{hassan1}. To calculate
$J_1(s)$ and $J_2(s)$ for $b>2a$, it is necessary to use Eq.
(\ref{init_as}), which describes RSA of small particles onto a
short line $x<b$. Its analytical solution and therefore the
explicit expressions for $J_1(s)$ and $J_2(s)$ can be obtained for
the case $b/a <5 $ using direct recursion to find $n_a(x)$  (for
one-particle RSA problem!). The result is rather cumbersome but
suitable for numerical integration.

For the case $b/a > 5 $ one can exploit the exponentially rapid
approach of the solution of Eq. (\ref{init_as}) to its asymptotic
behavior in the limit $x\gg1$ (see e.g. \cite{Lal} for the
numerical data). This asymptotic solution, \be
n_a(x)=\frac{R}{a}(x+1)-1 , \label{asymp_na} \ee can be used to
calculate $J_1(s)$ and then $J_2(s)$. To do this, we multiply Eq.
(\ref{init_as}) by $\exp(-sx)$ and integrate between 0 and $b-1$.
We obtain an equation for $J_2(s)$ of the form \ba
 J_2'(s)+\left(a+\frac{2}{s}e^{-as}\right)J_2(s)
=-\frac{1}{ s^2}\left\{e^{-as}I(s)\right.\nonumber
\\ \left.+s(b-a)e^{-bs}[n_a(b)-1]+2se^{-as}J_1(s) \right\}, \label{eq-J2}
\ea where \be I(s)=\int_0^{(b-a)s}dyye^{-y}. \label{eq-G_a}\ee
Solution of Eq. (\ref{eq-J2}), satisfying the boundary conditions
for $n_a$ given by Eq. (\ref{init_a}), is of the form \ba
J_2(s)&=&\frac{1}{ \tilde \beta(s) s^2}e^{-as}\int_0^s dt \tilde
\beta(t)\left[n_a(b)(b-a)te^{-(b-a)t}\right.\nonumber
\\&&\left.+2tJ_1(t)-\left(1-e^{-(b-a)t}\right)\right] \ea  with \be \tilde
\beta(t)=\exp \left[-2\int_0^{at} \left(
\frac{1-e^v}{v}dv\right)\right]. \ee The contribution of small
particles to the fill factor is then given by
\be\alpha_a=\int_0^\infty \beta(u)R_a(u)du \label{alph_a}.\ee in
which $\beta(u)$ is given by the Eq. (\ref{beta}) and $R_a(u)$ is
defined by Eq. (\ref{R a}). The obtained solution, though rather
unwieldy, is suitable for numerical integration and for $b/a>5$ it
gives the results that agree with Monte Carlo simulations.

In the limiting case $b/a=b/a\gg1$ it reduces to a more compact
final expression for the contribution to the total coverage from
the small particles \be \theta_a=R\left\{1+\int_0^\infty du
e^{-qu}\beta_p(u)\left[q(u-1)-2pe^{-u}\right] \right\},
\label{poin-smco}\ee with $\beta_p(u)$  defined by (\ref
{beta_p}). For $q=1$, Eq. (\ref{poin-smco}) properly gives
$\theta_a=R$, while for $q=0$ one has $\theta_a=R(1-R)$. The
latter expression corresponds to the coverage by small particles
of the gaps between the large particles left after their initial
deposition. For arbitrary $q$, the coverage given by Eq.
(\ref{poin-smco}) is depicted in Fig. 2 by the open squares.

Similar approach can be used to calculate the small particle
coverage variance. However, for $b/a>2$ the equation for the
Laplace transform of $u_a(x)$ given by Eq. (\ref{1equ-va}),
including all contributions to $N_a(s)$, becomes rather
impractical. In the limiting case $b/a\gg1$, when fluctuations of
the large particle gaps dominate the variance of small-particle
coverage, one gets a more compact result shown in Fig. 5.


\begin{thebibliography}{99}
\bibitem{Rodg}G. J. Rodgers and Z. Tavassoli, Phys. Lett. A { \bf 246}, 252 (1998).
\bibitem{viot}D. Boyer, J. Talbot, G. Tarjus, P. Van Tassel, and P. Viot,
 Phys. Rev. E {\bf 49}, 5525 (1994).
\bibitem{assl}A. V. Subashiev and S. Luryi,  Phys. Rev. E {\bf 75}, 011123 (2007).
\bibitem{bartelt}M. C. Bartelt and V. Privman,   Phys. Rev. A
{\bf 44}, R2227-R2230 (1991).
\bibitem{hassan1}M. K. Hassan, J. Schmidt, B. Blasius, J. Kurths,
 Phys. Rev. E {\bf 65}, 045103(R) (2002).
\bibitem{hassan2}M. K. Hassan and J. Kurths, J. Phys. A
{\bf 34}, 7517 (2001).
\bibitem{Araujo}N. A. M.  Araujo,    and A. Cadilhe,
 Phys. Rev. E {\bf 73}, 051602 (2006).
\bibitem{mao}D. J. Burridge and Y. Mao,  Phys. Rev. E { \bf 69}, 037102 (2004).
\bibitem{coff} E. G. Coffman, Jr., L. Flatto, P. Jelenkovich, and  B. Poonen,
 Algorithmica {\bf 22}, 448
(1998).
\bibitem{ino}M. Inoue, Phys. Rev. B {\bf 25}, 3856 (1982).
\bibitem{crack} P. Calka, A. Mezin, P. Vallois,
 Stochastic Processes and their Applications, {\bf 115},
983-1016 (2005).
\bibitem{Talbot}J. Talbot, G. Tarjus, P.R Van Tassel, P. Viot,
 Colloids and surfaces A:
Physicochemical and Engineering Aspects {\bf B 165}, 278 (2000).
\bibitem{Privman} V. Privman,
Colloids Surf A  {\bf 165} 231-240 (2000).
\bibitem{Devan}R. Devanathan,  L. R. Corrales, F. Gao, W. J. Weber,
 Nuclear
Instrum. Methods Phys. Res. A {\bf 565}, 637-649 (2006).
\bibitem{Spieler}H. Spieler, {\em Semiconductor Detector Systems},
Oxford University Press, 2005.
\bibitem{Klein}C. Klein,  J. Appl. Phys. {\bf 39}, 2029 (1968).
\bibitem{rusb}W. van Roosbroeck,  Phys. Rev. {\bf 139}, A 1702 (1965).
\bibitem{corr} This correlation originates from the basic fact that
a simple random division of a segment in two parts produces highly
correlated pieces: if one is short the other is long and vice versa.
Energy branching by impact ionization evidently has the similar
property, as the sum of secondary-particle energies is fixed by
energy conservation. This type of correlations was first pointed out
by Ugo Fano in 1947 \cite{Fano} and bears his name.
\bibitem{Fano}U. Fano,  Phys. Rev, {\bf 72}, 26 (1947).
\bibitem{Variat} N. H. Bingham, C. M. Goldie, J. L. Teugels,
{\em Regular Variation}, Cambridge University Press, Cambridge,
1987.
\bibitem{renyi}A. R\'enyi,  Publ. Math. Inst.
Hung. Acad. Sci. {\bf 3} 109 (1958); Trans. Math. Stat. Prob. {\bf 4}, 205 (1963).
\bibitem{Mack} J. K. Mackenzie,
 Journ. Chem. Phuys. {\bf 37}, 723 (1962).
\bibitem{asslr} S. Luryi and A. V. Subashiev,
unpublished.
\bibitem{Alkhas} G.D.Alkhazov, A.A. Vorob'ev, A.P Komar,
 Nucl. Instr. Meth. {\bf 48}, 1-12 (1967).
\bibitem{Nay}P. E. Nay, Annals of Mathematical Statistics, {\bf 33}, 702-718 (1962).
\bibitem{Kogan}Sh. Kogan, {\em Electronic Noise and Fluctuations in
Solids}, Cambridge University Press, Cambridge, 1996.
\bibitem{Lal} M. Lal and P. Gillard,  Math. Computation {\bf 28},
562 (1974).
\end{thebibliography}
\end{document}